\documentclass[debug,overfull,info]{epl}
\usepackage{amsmath}
\usepackage{epsf}
\usepackage{epsf}
\title{
Universal crossing probability in anisotropic systems
}
\author{L. Turban}
\institute{ 
  Laboratoire de Physique des Mat\'eriaux, UMR CNRS 7556, 
  Universit\'e Henri Poincar\'e (Nancy 1) 
  - BP 239, 54506 Vand\oe uvre l\`es Nancy Cedex, France
}


\begin{document}

\maketitle

\begin{abstract}
Scale-invariant universal crossing probabilities are studied for critical 
anisotropic systems in two dimensions. For weakly anisotropic standard 
percolation in a rectangular-shaped system, Cardy's exact formula is 
generalized using a length-rescaling procedure. For strongly anisotropic 
systems in $1+1$ dimensions, exact results are obtained for the random walk 
with absorbing boundary conditions, which can be considered as a linearized 
mean-field approximation for directed percolation. The bond and site 
directed percolation problem is itself studied numerically via Monte Carlo 
simulations on the diagonal square lattice with either free or periodic 
boundary conditions. A scale-invariant critical crossing probability is 
still obtained, which is a universal function of the effective aspect ratio 
$r_\ab{eff}=c\, r$ where $r=L/t^z$, $z$ is the dynamical exponent and $c$ is 
a non-universal amplitude. 
\end{abstract}

\section{Introduction}
The study of {\it universal crossing probabilities} in the standard 
random percolation problem at criticality has been the subject of 
much 
interest over the last decade~\cite{langlands92,cardy92,cross-perco,%
rigorous,span-clus-mc,span-clus-ex}
(see ref.~\cite{cardy01} for a review).
At the percolation threshold, $p_\ab{c}$, on a rectangular lattice 
with length $L_\parallel$ and width $L_\perp$, the probability to have 
at least one cluster crossing the system in one of the two directions 
(or in both) is a {\it scale-invariant} quantity in the {\it continuum 
limit} where the lattice spacing $a\to 0$. Actually, the crossing 
probability is a function $\pi(r)$ of the aspect ratio, 
$r=L_\perp/L_\parallel$, and depends also on the boundary conditions. It 
remains universal in the sense that it does not depend on the lattice 
type or whether one is considering site, bond or site-bond percolation. 

Following the pioneering numerical work of Langlands 
\etal~\cite{langlands92}, Cardy~\cite{cardy92} obtained an analytical 
expression for the crossing probability between opposite sides of a 
rectangle. He combined the correspondence between percolation and the 
limit $q\to1$ of the $q$-state Potts model~\cite{kasteleyn69} with 
boundary conformal field theory techniques to calculate the crossing 
probability between two segments on the boundary of a half-plane at 
criticality. A conformal mapping could then be used to transform the 
half-plane result into the rectangular geometry. Cardy's and another 
crossing probability formula have been recently proven 
rigorously~\cite{rigorous}.
Monte Carlo~\cite{span-clus-mc} and exact results~\cite{span-clus-ex} 
have been obtained for the probability to find $n$ incipient spanning 
clusters connecting two disjoint segments at the boundary of a finite 
system. More recently, crossing probabilities on same-spin Ising 
clusters in two dimensions have been also 
considered~\cite{cross-ising}. 

Let us notice that the existence of a {\it scale-invariant} crossing probability 
is not a trivial result. In the case of ordinary percolation, for example, it is 
linked to the vanishing, in the limit $q\to1$, of the scaling dimension $x(q)$ 
of a boundary condition changing operator of the Potts model~\cite{cardy92}.

In this letter, we investigate the behaviour of crossing 
probabilities in anisotropic critical systems. We first give an exact treatment 
of weakly anisotropic standard percolation. We show how a length rescaling 
technique, which has been used in other contexts~\cite{barber84}, allows to 
generalize Cardy's exact results in this situation.
Then we turn to crossing probabilities in strongly anisotropic systems. To our 
knowledge, such problems have not been considered before, at least with the 
appropriate anisotropic scaling analysis. We start with the random walk with 
absorbing boundary conditions in the continuum limit which can be considered as 
a linearized mean-field approximation~\cite{kaiser95} for directed 
percolation~\cite{broadbent57,kinzel83,hinrichsen00} and illustrates on 
an exactly solvable example the behaviour of the crossing probability in 
strongly anisotropic systems. Finally we present the results of Monte Carlo 
simulations for the crossing probability in directed percolation, a 
strongly anisotropic problem which has so far defied all attempts
at an exact solution, even in two dimensions. 

\section{Weakly anisotropic percolation}
We consider the random bond percolation on a square lattice with 
anisotropic bond occupation probabilities. The system is rectangular and 
the bond probability is $p_\parallel$ ($p_\perp$) in the direction 
parallel to the side with length $L_\parallel$ ($L_\perp$). 
The correlation lengths, $\xi_\parallel=\hat{\xi}_\parallel t^{-\nu}$, 
$\xi_\perp=\hat{\xi}_\perp t^{-\nu}$, diverge along the critical 
line where $t=0$ with the same exponent $\nu$ for both directions but 
different amplitudes $\hat{\xi}_\parallel$ and $\hat{\xi}_\perp$. The 
isotropy can be restored~\cite{barber84} by rescaling the lattice 
parameter anisotropically so that 
$\xi_\parallel a_\parallel=\xi_\perp a_\perp$. The aspect ratio $r$ 
is then changed into an effective one 
\begin{equation}
r_\ab{eff}=\frac{L_\perp a_\perp}{L_\parallel a_\parallel}
=\frac{L_\perp/\xi_\perp}{L_\parallel/\xi_\parallel}
=\frac{\hat{\xi}_\parallel}{\hat{\xi}_\perp}r\,.
\label{e-reff}
\end{equation}
In the second expression the effective aspect ratio involves the 
lengths measured in units of the corresponding correlation lengths.

The critical amplitude ratio 
$c=\hat{\xi}_\parallel/\hat{\xi}_\perp$, which is non-universal, 
is known for the $q$-state Potts model on the square 
lattice~\cite{kim87}. In the percolation limit, $q\to1$, one obtains:
\begin{equation}
c=\tan\frac{3 u}{2}\,,\qquad \frac{\sin u}{\sin(\pi/3-u)}
=\frac{p_{\ab{c}\parallel}}{1-p_{\ab{c}\parallel}}
=\frac{1-p_{\ab{c}\perp}}{p_{\ab{c}\perp}}\,,
\label{e-rescal}
\end{equation}
which solves the problem for a weak anisotropy. When the anisotropy 
axis is not parallel to one edge of the rectangle, the rescaling 
leads to an isotropic system with the shape of a parallelogram.

\section{Random walk}
\begin{figure}
\onefigure[width=7cm]{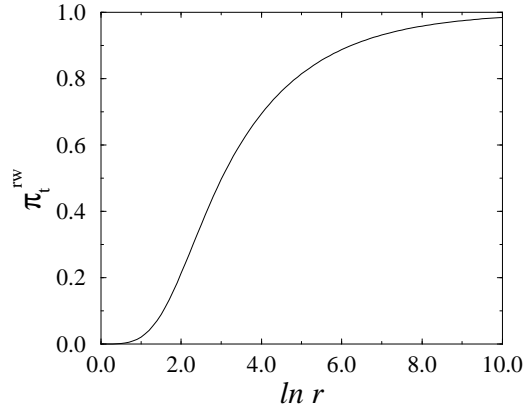}
\caption{Crossing probability in the time direction for a random walk 
with  
absorbing boundary conditions at $l=0$ and $L$ and a uniform initial 
probability density at $t=0$. $r=L^2/t$ is the scale-invariant aspect 
ratio 
of the system and $D=1$.}
\label{f.1}
\end{figure}

Standard conformal methods do not apply in strongly anisotropic systems.
The simplest example of such a system is the random 
walk in $1+1$~dimensions, which may be considered as a directed walk 
in space-time. In the continuum limit, the probability density 
$P(x,t)$ 
satisfies the diffusion equation
\begin{equation}
\frac{\partial P}{\partial t}=D\frac{\partial^2 P}{\partial x^2}\,.  
\label{e-dif}
\end{equation}
We assume a uniform initial probability density $P(x,0)=1/L$ and 
absorbing boundary conditions at~$x=0$ and~$L$. The solution is then 
obtained as the following eigenvalue expansion:
\begin{equation}
P(x,t)=\sum_{p=0}^\infty\frac{4}{(2p+1)\pi L}
\sin\left[(2p+1)\frac{\pi x}{L}\right]
\exp\left[-\frac{(2p+1)^2\pi^2Dt}{L^2}\right]\,.
\label{e-P}
\end{equation}
On the rectangle $L\times t$, the crossing probability in the time 
direction is 
given by:
\begin{equation}
\pi^\ab{rw}_t=\int_0^LP(x,t)\,\upd x
=\sum_{p=0}^\infty\frac{8}{(2p+1)^2\pi^2}
\exp\left[-\frac{(2p+1)^2\pi^2Dt}{L^2}\right]\,.
\label{e-cross-rw}
\end{equation}
Thus the crossing probability shown in fig.~\ref{f.1} is a universal 
function of the scaled variable $cL^z/t$, product of the aspect ratio 
$r=L^z/t$, where $z=2$ is the dynamical 
exponent for the random walk, by the non-universal inverse diffusion 
constant $c=D^{-1}$. 

In strongly anisotropic systems, such a behaviour generally follows 
from anisotropic scaling~\cite{binder89}. Under the anisotropic scale 
transformation, $L'=L/b$, $t'=t/b^z$, the crossing probability 
behaves as:
\begin{equation}
\pi_t(L,t)=b^{-x_{\pi_t}}\,\pi_t\left(\frac{L}{b},\frac{t}{b^z}\right
)
\,.
\label{e-scaling}
\end{equation}
For a dimensionless crossing probability, $x_{\pi_t}=0$, one obtains  
the scale-invariant behaviour mentioned above with $b=L$.

The probability density for a walker starting at $t=0$ with uniform 
probability on the segment $[0,L]$ to die at time $t$ is given by
\begin{equation}
-\frac{\upd \pi^\ab{rw}_t}{\upd t}
=\frac{8D}{L^2}\sum_{p=0}^\infty 
\exp\left[-\frac{(2p+1)^2\pi^2Dt}{L^2}\right]
=\frac{4D}{L^2}\,\vartheta_2\left[
\exp\left(-\frac{4\pi^2Dt}{L^2}\right)\right]\,,
\label{e-surv-rw}
\end{equation}
where $\vartheta_2(u)=\sum_{p=-\infty}^{+\infty}u^{(p+1/2)^2}$ is a 
Jacobi theta function.

Since the linearized mean-field approximation for directed 
percolation~\cite{kinzel83,kaiser95} is also
governed by the diffusion equation~(\ref{e-dif}),
the mean-field crossing probability for directed percolation is given 
by~(\ref{e-cross-rw}) when the aspect ratio is sufficiently small.

\section{Directed percolation}
The critical crossing probability in the time direction, 
$\pi^\ab{dp}_t$, 
has been studied via Monte Carlo simulations for bond and site 
directed percolation on the square lattice, with the time axis along the 
diagonal direction. Space and time coordinates take integer and 
half-integer values alternatively on successive spatial rows and 
$t\geq0$. There are two directed bonds leaving a site at $(x,t)$ and 
terminating on the sites at $(x\pm1/2,t+1/2)$. 

Sites may be wet or dry and a directed percolation cluster is a 
collection of connected wet sites starting from some source at 
$t=0$~\cite{broadbent57,kinzel83,hinrichsen00}. In the bond problem, bonds 
are open with probability $p$. A site is wet at time $t+1/2$ when it is 
connected through an open bond to a site which was wet at time $t$. In the 
site problem, sites are occupied with probability $p$ and all the bonds 
are open. A site is wet at time $t+1/2$ when it is occupied and connected 
to a site which was wet at time $t$. A {\it scale-invariant crossing 
probability} is obtained when all the sites are wet in the initial 
state at $t=0$.

\begin{figure}
\twoimages[width=7cm]{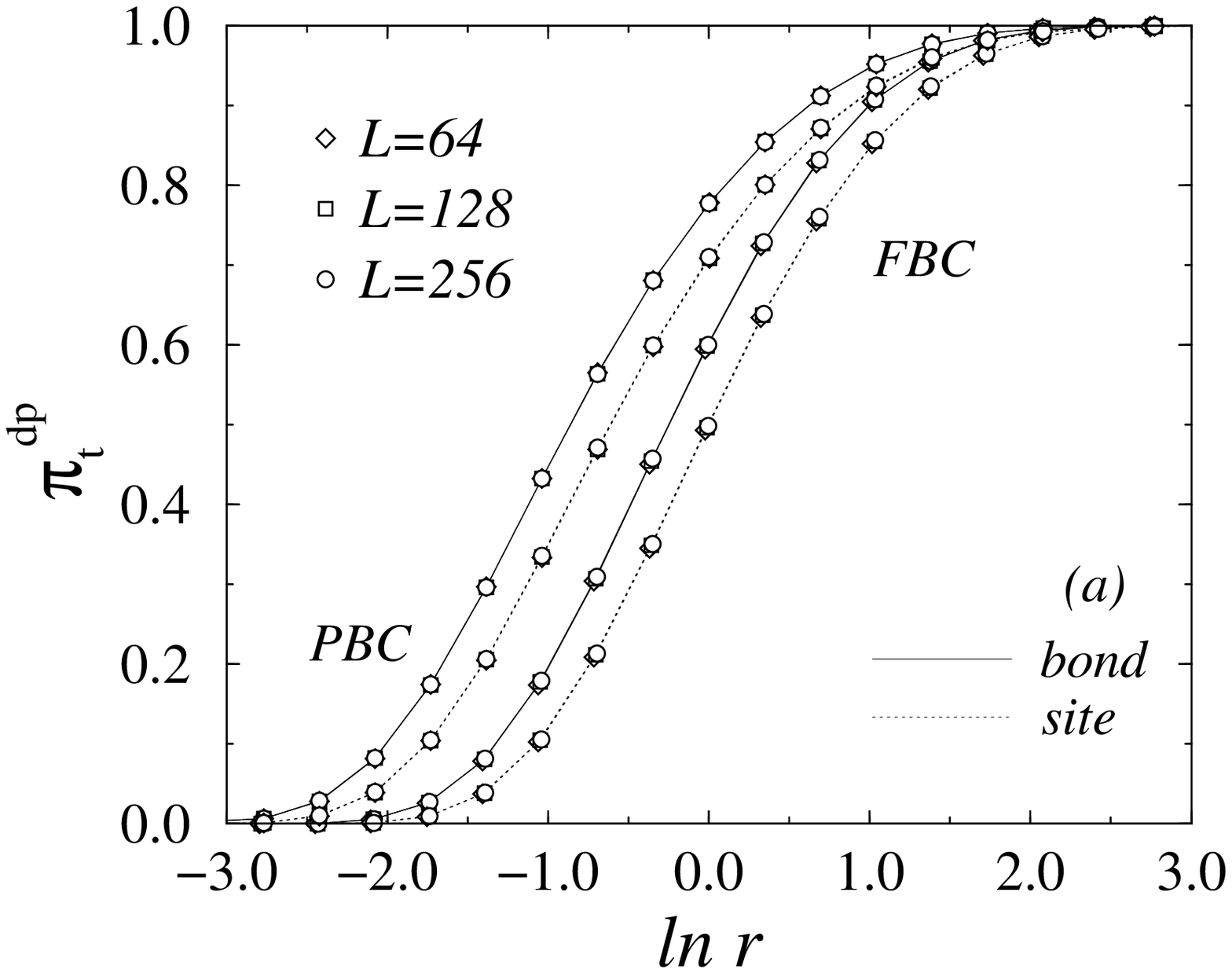}{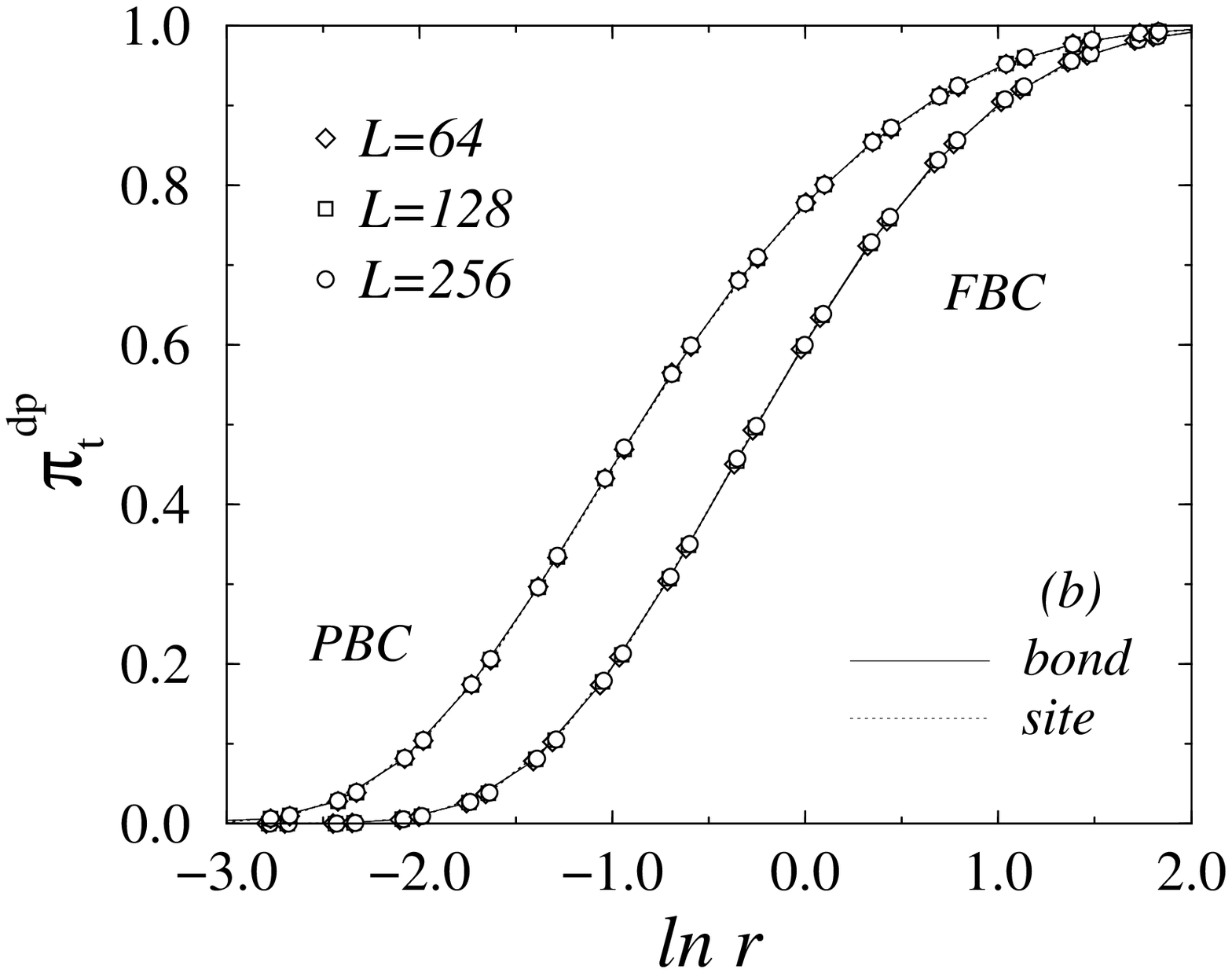}
\caption{(a) Crossing probability $\pi^\ab{dp}_t$ in the time 
direction 
as a function of the logarithm of the aspect ratio $r=L^z/t$ for site 
and 
bond directed percolation on the diagonal square lattice with either 
periodic or free boundary conditions in the space direction. The $L$ 
sites are wet at $t=0$. (b) Shifting the curves for the site problems 
by 
the same amount, the data collapse on a single scale-invariant 
universal 
curve for each type of boundary conditions.}
\label{f.2ab}
\end{figure}

The critical crossing probability was calculated for $10^6$ samples 
with 
sizes $L=2^3$ to $2^8$ and different values of the aspect ratio, 
$r=L^z/t$, ranging from $r\simeq 2^{-5}$ to $2^5$. The percolation 
thresholds $p_\ab{c}^\ab{bond}=0.644700185(5)$, 
$p_\ab{c}^\ab{site}=0.70548522(4)$ and the dynamical exponent 
$z=1.580745(10)$ were taken from ref.~\cite{jansen99}. Either 
periodic 
boundary conditions (PBC) or free boundary conditions (FBC) were used 
at 
$l=1$ and $L$. The samples were grouped into 20 packets in order to 
evaluate the errors on $\pi^\ab{dp}_t$. The raw data $\pi^\ab{dp}_t$ 
versus $\ln r$ are shown in fig.~2a for the largest sizes. The 
statistical 
errors are actually smaller than the size of the symbols. The scale 
invariance is manifest and finite size corrections to the leading 
behaviour are quite weak. There is a shift in $\ln r$ between the 
curves 
for site and bond percolation which is the same for PBC and FBC. As 
shown 
in fig.~2b, an universal curve is obtained for each type of boundary 
condition after the curves corresponding to the site problem have 
been 
shifted by the same amount.

In this strongly anisotropic problem, the correlation lengths, 
$\xi_\perp$ in the space direction and $\xi_\parallel$ in the time 
direction, diverge as 
\begin{equation}
\xi_\perp=\hat{\xi}_\perp \vert p-p_\ab{c}\vert^{-\nu}\;,
\qquad \xi_\parallel=\hat{\xi}_\parallel \vert 
p-p_\ab{c}\vert^{-z\nu}\;.
\label{e-xi}
\end{equation}
Thus, generalizing the behaviour for the weakly anisotropic system in 
eq.~(\ref{e-reff}), for each type of boundary conditions, $i=FBC$ or 
$PBC$, one expects the crossing probability $\pi^\ab{dp}_t$ to be an 
universal function $f_i(r_\ab{eff})$ of the scale-invariant ratio
\begin{equation}
r_\ab{eff}=\frac{(L/\xi_\perp)^z}{t/\xi_\parallel}=c\, r\;,
\qquad c=\frac{\hat{\xi}_\parallel}{(\hat{\xi}_\perp)^z}\;,
\label{e-rlt}
\end{equation}
in agreement with anisotropic scaling in eq.~(\ref{e-scaling}).
As above for weakly anisotropic percolation and for the random-walk,   
the constant $c$ is non-universal and takes different values, 
$c_\ab{s}$ 
and $c_\ab{b}$, for the site and bond directed percolation on the 
diagonal square lattice. 

For a given type of boundary conditions, the crossing probability 
$\pi^\ab{dp}_t$ is the same for couples of values of the aspect ratio 
$r_\ab{s}$ and $r_\ab{b}$ such that the arguments of the universal 
function $f_i(r_\ab{eff})$ are the same for the site and bond 
problems. 
This happens when $c_\ab{s}r_\ab{s}=c_\ab{b}r_\ab{b}$. Thus, with a 
logarithmic scale, the two curves have a relative shift  
\begin{equation}
\ln r_\ab{s}-\ln 
r_\ab{b}=\ln\left(\frac{c_\ab{b}}{c_\ab{s}}\right)\;.
\label{e-shift}
\end{equation}

\begin{figure}
\onefigure[width=7cm]{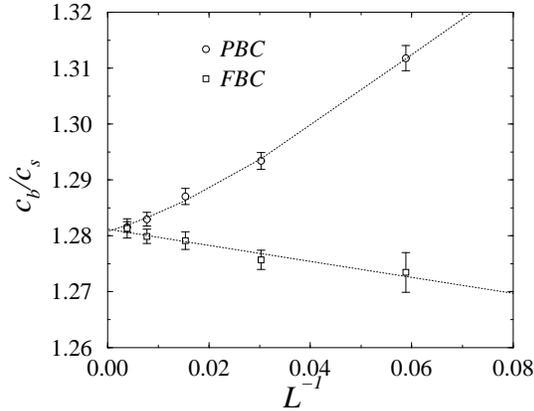}
\caption{Non-universal amplitude ratio $c_\ab{b}/c_\ab{s}$ as defined 
in 
eq.~(\protect\ref{e-rlt}) as a function of the inverse of the system 
size 
$L$. The dotted line corresponds to a cubic (quadratic) fit of the 
data 
for periodic (free) boundary conditions.}
\label{f.3}
\end{figure}

This shift was estimated for the different sizes for 6 points in the 
ascending part of the the curves. Since the points for the site and 
bond 
problems correspond to slightly different ordinates, a direct 
estimation 
of the shift was not possible. Instead we used a polynomial 
interpolation 
for the curve corresponding to the bond problem and the value of the 
abscissa associated with the common value of the ordinate was 
determined through the tangent method. The results for the ratio of 
non-universal amplitudes as a function of the inverse size are shown 
in 
fig.~\ref{f.3} for PBC and FBC. They converge from above and from 
below 
to the common asymptotic value
\begin{equation}
\frac{c_\ab{b}}{c_\ab{s}}=1.281(1)\;,
\label{e-amprat}
\end{equation}
corresponding to the shift $\ln r_\ab{s}-\ln r_\ab{b}=0.2476(8)$ 
which was 
used to obtain the data collapse in fig.~2b.
\section{Final remarks}
The same universal crossing probability is obtained for any finite 
homogeneous density of wet sites in the initial state at $t=0$ 
although the 
finite-size corrections to the leading scale-invariant behaviour is 
stronger at large values of $r$ (small $t$) when the initial density 
decreases.
With a single wet site at $l=L/2$ in the initial state, i.e., for an 
asymptotically vanishing initial density, the crossing probability 
remains a function of $r_\ab{eff}$ but the scale-invariance is lost, 
$\pi^\ab{dp}_t$ decaying as $L^{-\beta/\nu}$. The probability 
$\pi^\ab{dp}_t(n)$ to find $n$ clusters crossing the system in the time 
direction at $p_\ab{c}$ has been also studied. Details will be given elsewhere.

This work can be extended by considering the crossing probability on same-spin 
clusters in strongly anisotropic spin systems, like the three-dimensional 
uniaxial ANNNI model at the Lifshitz point, for which presumably exact results 
for the form of the correlation functions, following from a generalization of 
local scale invariance~\cite{henkel02}, have been obtained 
recently~\cite{pleimling01}.
 
\acknowledgments
The author wishes to thank {\scshape Lev Shchur} for a stimulating 
introductory talk about spanning clusters.

\end{document}